\documentclass[twocolumn,showpacs,showkeys,preprintnumbers,amsmath,amssymb]{revtex4}

\usepackage{graphicx}
\usepackage{dcolumn}
\usepackage{bm}
\usepackage[utf8x]{inputenc}
\usepackage{float}

\begin{document}


\title{Cornell Potential: A Neural Network Approach} 

\author{Halil Mutuk}
\affiliation{%
Physics Department, Faculty of Arts and Sciences, Ondokuz Mayis University, 55200, Samsun, Turkey
}%


\begin{abstract}
In this study, we solved Schrödinger equation with Cornell potential (Coulomb-plus-linear potential) by using neural network approach. Four different cases of Cornell Potential for different potential parameters were used without a physical relevance. Besides that charmonium, bottomonium and bottom-charmed spin-averaged spectra were also calculated. Obtained results are in good agreement with the reference studies and available experimental data. 
\end{abstract}

\pacs{12.39.Pn, 87.18.Sn}
\keywords{Cornell potential, Charmonium, Bottomonium, Bottom-Charmed Mesons, Neural Networks}
\maketitle

\section{\label{sec:level1}Introduction}
The Schrödinger equation with Coulomb-plus-linear potential (Cornell potential) has received a great deal of attention  as an important non-relativistic model for the study of quark-antiquark systems, namely mesons \cite{1,2,3,4,5,6,7,8,9,10,11,12,13,14,15,16,17,18}. Especially Coulomb plus linear potential was the first potential model to study heavy quarkonium systems and inspired other phenomenological models.  The quarkonium spectrum is a substantial field to improve our understanding about nature of QCD. The Cornell potential reads as
\begin{equation}
V(r)=-\frac{a}{r}+br,
\end{equation}
where $a$ and $b$ are some constants. At small quark-antiquark distances, the Cornell potential goes like $\approx \frac{1}{r}$ which is known as Coulomb-like term and at large distances it goes as $\approx r$, which is known as linear term. The Coulomb-like term arises from one-gluon exchange and linear term presumably arises from higher order effects \cite{19}. Upto now, the linear term has not been calculated from the first principles of QCD. 

Aside from its physical relevance, the Schrödinger equation with Coulomb-plus-linear  potential have been studied with pure mathematical techniques. Hall used the method of potential envelopes to construct
general upper and lower bounds on the eigenvalues of the Hamiltonian representing a single particle in
Coulomb plus linear potential \cite{20}. In \cite{21}, Chaudhuri et al. used Hill determinant method to bound state eigenvalue problem with Coulomb-plus-linear potential. Plante and Antippa solved the Schrödinger equation for a quark–antiquark system interacting via a Coulomb-plus-linear potential, and obtained the wave functions as power series, with their coefficients given in terms of the combinatorics functions \cite{22}. In \cite{23}, the authors presented a numerically precise treatment of the Crank–Nicolson method with an imaginary time evolution operator in order to solve the Schrödinger equation.  Although such models have been studied to solve Schrödinger equations numerically or analytically, exact solutions of Schrödinger equation with Cornell potential are still unknown. 

Artificial Neural Networks (ANNs) can be used as an alternative method to solve both ordinary and partial differential equations. ANN is a parallel distributed processor in which numerous numbers of simply designed computing units exist. These units are called \textit{neurons}. Its massive connections between neurons can be used for storing various types of informations, particularly the ones classified as \textit{knowledge} or \textit{experience} \cite{24,25}. These informations are acquired as interneuron connection strengths, or synaptic weights.

Especially with the work of \citep{26,27}, ANNs are being used widely for solving ordinary and partial differential equations. ANNs have many advantages compared to existing semi-analytic and numerical methods. The main advantages of using neural networks to solve differential equations can be stated as follows \cite{28}:
\begin{itemize}
\item Neural networks (NNs) are universal functions approximators which are useful in solving differential equations.
\item Trial solutions of artificial neural networks involve  a single independent variable regardless of the dimension of the problem.
\item The approximate solutions are continuous over all the domain of integration. In contrast, the numerical methods provide solutions only over discrete points and the solution between these points must be interpolated.
\item The computational complexity does not increase considerably with the number of sampling points and with the number of dimensions in problem.
\end{itemize}

Being an eigenvalue problem, Schrödinger equation had got attention by ANNs at the beginning of 90s.
In \cite{29}, the authors teached a neural network the solution of the two dimensional Schrödinger equation for some model potential energy functions. In work of Androsiuk et al. \cite{30}, they presented computer simulations of a neural network capable of learning to perform transformations generated by the Schrödinger equation required to find eigenenergies of two dimensional harmonic oscillator. Sugawara  presented a new approach for solving the Schrödinger equation based on genetic algorithm and artificial neural network. The method was tested in the calculation of one dimensional harmonic oscillator and other model potentials such as Morse potential \cite{31}. 

In the present paper, we solve Schrödinger equation with Coulomb-plus-linear potential via ANN system. In Section \ref{sec:level2}, we introduce the formalism of ANN method and describe how it can be applied to the quantum mechanical calculations. In Section \ref{sec:level3}, we give and discuss numerical results for the eigenvalues of Coulomb-plus-linear potential and in Section \ref{sec:level4} we summarize our findings. 

\section{\label{sec:level2}Formalism}
\begin{figure}[H]
\includegraphics[width=3.4in]{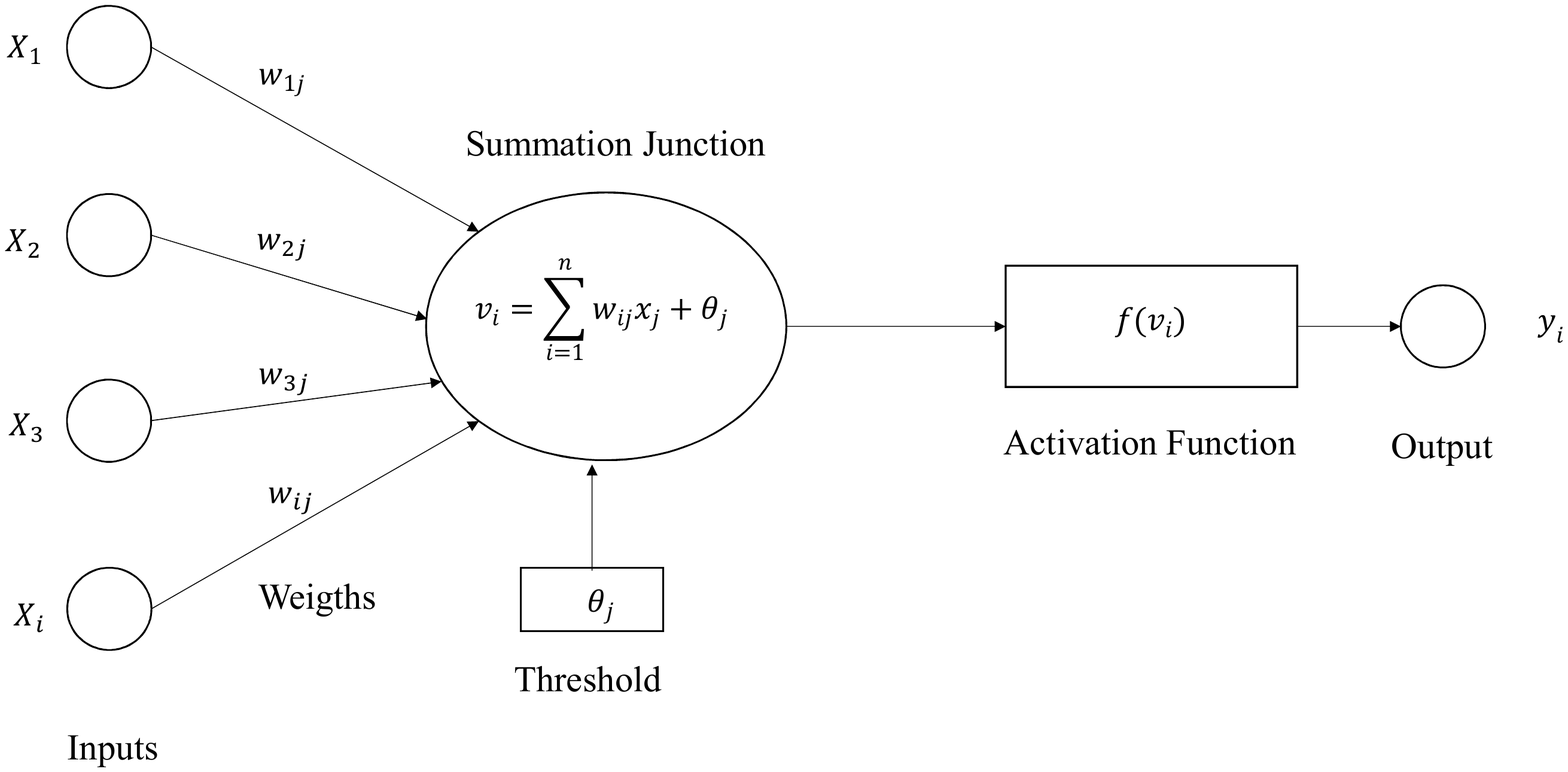}
\caption{\label{fig:1} A model of single neuron}
\end{figure}

The neural network (NN) is constructed as a model of simply designed computing unit, called {\it neuron}. Fig. \ref{fig:1} illustrates a simple model of a single neuron with multiple inputs and one output.

\begin{figure}[H]
\includegraphics[width=3.4in]{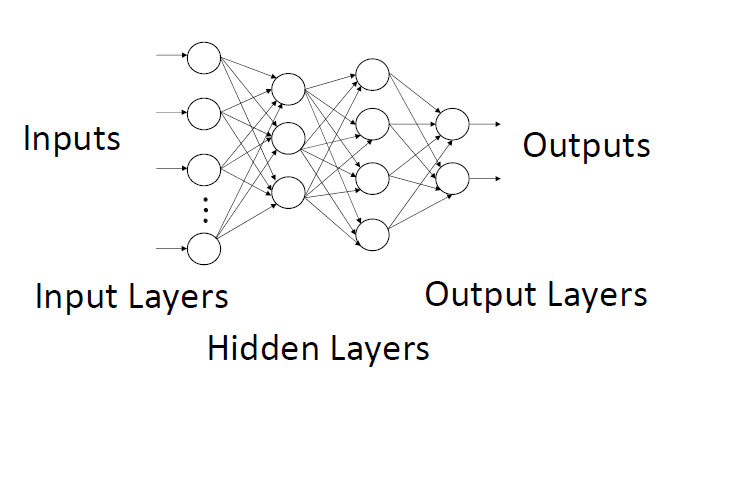}
\caption{\label{fig:2} A model of multilayer neural networks}
\end{figure}

All the input signals are summed up as $z$ and the output signal is determined by the nonlinear activation function $\sigma(z)$. In this work we use a sigmoid function
\begin{equation}
\sigma(z)=\frac{1}{1+e^{-z}} \label{sigmoid}
\end{equation}
as an activation function since it is possible to derive all the derivatives of $\sigma(z)$ in terms of itself. This differentiability is an important aspect for the Schrödinger equation. Here we use a multilayer perceptron neural network (MLPN). The architecture of MLPN is shown in Fig. \ref{fig:2}.

This is an example of feed forward neural network. Feed forward neural networks are the most popular architectures due to their structural flexibility, good representational capabilities and availability of a large number of training algorithm \cite{24}. This network consists of neurons arranged in layers in which every neuron is connected to all neurons of the next layer and the information process can be in only one way, from input to the output. The input-output properties of the neurons (perceptrons in computerized systems) can be written as 
\begin{eqnarray}
o_i&=&\sigma(n_i), \\
o_j&=&\sigma(n_j), \\
o_k&=&\sigma(n_k),
\end{eqnarray}
where $i$ is for input, $j$ is for hidden and $k$ is for output layers. Input to the perceptrons are given as
\begin{eqnarray}
n_i&=&(\text{Input signal to the NN}), \\
n_j&=& \sum_{i=1}^{N_i} \omega_{ij}o_i+ \theta_j, \\
n_k&=& \sum_{i=1}^{N_j} \omega_{jk}o_j+ \theta_k.
\end{eqnarray}
Here, $N_i$ and $N_j$ represents the numbers of the units belonging to input and hidden layers, $\omega_{ij}$ is the synaptic weight parameter which connects the neurons $i$ and $j$ and $\theta_j$ represents threshold parameter for the neuron $j$ \cite{30}. The overall response of the network is given as
\begin{equation}
o_k=\sum_{j=1}^{b_n} \omega_{jk}\sigma \left( \sum_{i=1}^{a_n}\omega_{ij}n_i +\theta_j \right)+ \theta_k. \label{eqn1}
\end{equation}
The derivatives of $o_k$ with respect to network parameters (weights and thresholds) can be done by differentiating Eqn. (\ref{eqn1}) as
\begin{eqnarray}
\frac{\partial o_k}{\partial \omega_{ij} }&=& \omega_{jk} \sigma^{(1)}(n_j)n_i,\\
\frac{\partial o_k}{\partial \omega_{jk} }&=& \sigma(n_j)\delta_{kk^\prime},\\
\frac{\partial o_k}{\partial \theta_j }&=& \omega_{jk} \sigma^{(1)}(n_j),\\
\frac{\partial o_k}{\partial \theta_k }&=&\delta_{kk^\prime}.
\end{eqnarray}
Neural network systems are self-training systems. Therefore parameters of training need updating in the transfer process and these equations play important roles on the learning processes of the neural networks.

\subsection{Implementation of ANN to Quantum System}
We consider the application of the ANN to a quantum mechanical calculation. We will follow the formalism which was developed in \cite{26}. Consider the following differential equation
\begin{equation}
H\Psi(r)=f(r) \label{eqn2}
\end{equation}
where $H$ is a linear operator, $f(r)$ is a known function and  $\Psi(r)=0$ at the boundaries. In order to solve this differential equation a trial function can be written as
\begin{equation}
\Psi_t(\textbf{r})=A(\textbf{r})+B(\textbf{r}, \textbf{$\lambda$})N(\textbf{r}, \textbf{p}),
\end{equation}
which employs a feed forward neural network with parameter vector $\textbf{p}$ and $\textbf{$\lambda$}$ to be adjusted. The parameter vector $\textbf{p}$ corresponds to the weights and biases of the neural architecture. The functions $A(\textbf{r})$ and $B(\textbf{r}, \textbf{$\lambda$})$ should be specified in a convenient way so that $\Psi_t(\textbf{r})$ satisfies the boundary conditions regardless of the $\textbf{p}$ and $\textbf{$\lambda$}$ values.  To obtain a solution for Eqn. (\ref{eqn2}), the collocation method can be used and the differential equation can be transformed into a minimization problem 
\begin{equation}
\underset{p,\lambda}{\min} \sum_i \left[ H\Psi_t(r_i)-f(r_i)  \right]^2. \label{dif2}
\end{equation}
For Schrödinger equation, Eqn. (\ref{eqn2}) takes the form
\begin{equation}
H\Psi(r)=\epsilon \Psi(r) 
\end{equation}
with the boundary condition $\Psi(r)=0$. In this case, the trial solution can be written as
\begin{equation}
\Psi_t(r)=B(\textbf{r}, \textbf{$\lambda$})N(\textbf{r}, \textbf{p})
\end{equation}
where $B(\textbf{r}, \textbf{$\lambda$}) =0$ at boundary conditions for a range of $\lambda$ values. By discretizing the domain of the problem, it is transformed into a minimization problem with respect to the parameters $\textbf{p}$ and $\textbf{$\lambda$}$
\begin{equation}
E(\textbf{p},\textbf{$\lambda$})=\frac{\sum_i \left[ H\Psi_t(r_i, \textbf{p},\textbf{$\lambda$})-\epsilon \Psi_t(r_i, \textbf{p},\textbf{$\lambda$})  \right]^2}{\int \vert \Psi_t \vert^2 d\textbf{r}},
\end{equation}
where $E$ is the error function and $\epsilon$ can be computed as
\begin{equation}
\epsilon=\frac{\int  \Psi_t^{\ast} H \Psi_t  d\textbf{r}}{\int \vert \Psi_t \vert^2 d\textbf{r}}.
\end{equation}

Let's consider a multilayer perceptron with $n$ input units, one hidden layer with $m$ units and one output. This multilayer perceptron looks like Fig.\ref{fig:2} but with just one output. For a given input vector
\begin{equation}
\textbf{r}=\left( r_1, \cdots,r_n  \right),
\end{equation}
the output of the network is 
\begin{equation}
N=\sum_{i=1}^m \nu_i \sigma(z_i),
\end{equation}
where 
\begin{equation}
z_i=\sum_{j=1}^n \omega_{ij}r_j+u_i.
\end{equation}
Here, $\omega_{ij}$ denotes the weight from the input unit $j$ to the hidden unit $i$, $\nu_i$ is the weight from the hidden unit $i$ to the output, $u_i$ is the bias of hidden unit $i$ and $\sigma(z)$ is the sigmoid function, given in Eqn.(\ref{sigmoid}). The derivatives of network output can be defined as
\begin{equation}
\frac{\partial^k N}{\partial r^k_j}=\sum_{i=1}^m \nu_i \omega_{ij}^k \sigma_i^{(k)}
\end{equation}
where $\sigma_i=\sigma(z_i)$ and $\sigma^{(k)}$ is the $k$th order derivative of the sigmoid. 

Once the derivatives of the error with respect to the network parameters have been defined, any minimization technique can be carried out. In this work, we used a feed forward neural network with a back propagation algorithm. 

By employing this approach it is possible to obtain energy eigenvalues of the Cornell potential. Before obtaining Cornell potential eigenvalues, it will be useful to give an example.

\subsubsection{A Warm up example: Yukawa Potential}
The Yukawa potential have an important role in various branches of physics. For example, in plasma physics, it is known as the Debye-Hückel potential, in solid-state physics and atomic physics, it is called the Thomas-Fermi or the screened Coulomb potential and has a role in the nucleon-nucleon
interaction arising out of the one-pion-exchange mechanism in nuclear physics \cite{33}. The Yukawa potential reads as
\begin{equation}
V(r)=-\frac{A}{r}e^{-\alpha r}
\end{equation}
where the parameters $A$ and $\alpha$ are given by different expressions, depending on the problem under consideration. Yukawa potential does not admit an exact solution and therefore various approximate analytic and numerical methods. The ground state energy for $\hbar=m=1$, $A=\sqrt{2}$, $g=0.002$ in $\alpha=gA$ reported by \cite{33} is $E=-0.996006$, by \cite{34} is $E=-0.99601$ and by \cite{35} is $E=-0.9960$. 

We parametrize trial function as
\begin{equation}
\phi_t(x)=e^{-\beta x^2} N(x, \textbf{u}, \textbf{w}, \textbf{v}), ~ \beta >0
\end{equation}
where $N$ denotes a feed forward artificial neural network with one hidden layer and $m$ sigmoid hidden units
\begin{equation}
N(x, \textbf{u}, \textbf{w}, \textbf{v})=\sum_{j=1}^m \nu_j \sigma(\omega_j x+ u_j).
\end{equation}
The minimization problem is 
\begin{equation}
\frac{\sum_i \left[ H\phi_t(x_i)-\epsilon \phi_t(x_i)  \right]^2 }{\int \vert \phi_t(x) \vert ^2 dx}.
\end{equation}
The energy eigenvalue obtained with this scheme is $E=-0.996000170000301$ which is in good agreement with the numerical results in first five digits.

\section{\label{sec:level3}Numerical Results and Discussion}
 The radial part of the Schrödinger equation for Coulomb-plus-linear potential is
\begin{equation}
\frac{d^2 \phi(r)}{dr^2}+\frac{2m}{\hbar^2} \left( E- \frac{l(l+1)}{2mr^2} \hbar^2 +V(r)  \right) \phi(r)=0 \label{rad}
\end{equation}
where $V(r)=-\frac{a}{r}+br$ and $m=\hbar=1$. We trained the network with 200 equidistance points with $m=10$ in the interval $-6 \leq r \leq 6$ and solved the Schrödinger equation in four cases. 

\subsection{Case \textrm{1}: \textbf{$V(r)=-\frac{1}{r}+r$}}
Putting $V(r)=-\frac{1}{r}+r$ into the Eqn. (\ref{rad}) we get
\begin{equation}
-\frac{1}{2}\frac{d^2 \phi(r)}{dr^2}-\left( -\frac{1}{r}+r- \frac{l(l+1)}{2r^2}   \right) \phi(r)=E \phi(r) .
\end{equation}
The Hamiltonian of this equation is
\begin{equation}
H=-\frac{1}{2} \frac{d^2}{dr^2}-\left( -\frac{1}{r}+r- \frac{l(l+1)}{2r^2}   \right).
\end{equation}
By applying the procedure which was mentioned in the previous section, we get eigenvalues of this potential. In Tables \ref{tab:table1} and \ref{tab:table2}, we present eigenvalues of the potential $V(r)=-\frac{1}{r}+r$ for $n$ and $l$ states, respectively. 

\begin{table}[H]
\caption{\label{tab:table1}Comparison of eigenvalues of $V(r)=-\frac{1}{r}+r$ for $n$ states.}
\begin{ruledtabular}
\begin{tabular}{cccccccc}
 $l$&$n$&$E_{n0}$& \cite{13} &\cite{32}\\
\hline
0& 0 & $1.3978757858276367$ & $1.397875641660$ &$1.39788$ \\
 & 1 & $3.4750833511352539$ & $3.475086545396$ &$3.47509$ &\\
 & 2 & $5.0329142808914185$  & $5.032914359536$ &$5.03291$ &\\
 & 3 & $6.3701504468917847$ & $6.370149125486$ &$6.37015$ &\\
 & 4 & $7.5749331712722778$ & $7.57493264059~$ &$7.57493$ &\\
 & 5 & $8.6879128217697144$  & $8.687914590401$  &$8.68791$ &\\
\end{tabular}
\end{ruledtabular}
\end{table}

\begin{table}[H]
\caption{\label{tab:table2}Comparison of eigenvalues of $V(r)=-\frac{1}{r}+r$ for $l$ states.}
\begin{ruledtabular}
\begin{tabular}{cccccccc}
 $n$&$l$&$E_{0l}$& \cite{13} &\cite{32} \\
\hline
0& 0 & $1.3978757858276367$ & $1.397875641660$ &$1.39788$ &\\
 & 1 & $2.8256469964981079$ & $2.825646640704$ &$2.82565$ &\\
 & 2 & $3.8505786657333374$ & $3.850580006803$ &$3.85058$ &\\
 & 3 & $4.7267490625381470$ & $4.726752007096$ &$4.72675$ &\\
 & 4 & $5.5169814825057983$ & $5.516979644329$ &$5.51698$ &\\
 & 5 & $6.2483960390090942$ & $6.248395598411$ &$6.24840$ &\\
\end{tabular}
\end{ruledtabular}
\end{table}

It can be seen from Tables \ref{tab:table1} and \ref{tab:table2} that, eigenvalues of the radially and orbitally excited states by ANN method is in good agreement with the reference studies. 

\subsection{ Case \textrm{2}: $V(r)=-\frac{1}{r}+0.01r$}
The Hamiltonian for this potential is
\begin{equation}
H=-\frac{1}{2} \frac{d^2}{dr^2}-\left(-\frac{1}{r}+0.01r- \frac{l(l+1)}{2r^2}   \right).
\end{equation}
In Tables \ref{tab:table3} and \ref{tab:table4}, we present eigenvalues of the potential $V(r)=-\frac{1}{r}+0.01r$ for $n$ and $l$ states, respectively.

\begin{table}[H]
\caption{\label{tab:table3}Comparison of eigenvalues of $V(r)=-\frac{1}{r}+0.01r$ for $n$ states.}
\begin{ruledtabular}
\begin{tabular}{cccccccc}
 $l$&$n$&$E_{n0}$& \cite{13} &\cite{32} \\
\hline
0& 0 & $-0.221030285825463$ & $−0.221030563404$ &$-0.221031$\\
 & 1 & $0.0347223105947489$  & $0.034722241998$ &$0.0347222$\\
 & 2 & $0.1419130671269874$  & $0.141913022811$ &$0.141913$\\
 & 3 & $0.2202872073658741$ & $0.220287171811$ &$0.220287$\\
 & 4 & $0.2861103278041072$ & $0.344602792592$ $^{a}$ &$0.286111$\\
 & 5 & $0.3446028206061204$  & $0.448055673514$ $^{a}$ &$0.344602$ \\

\end{tabular}
\end{ruledtabular}
$^{a}$ In a private communication with R. L. Hall, corresponding author of \cite{13}, he mentioned there was a copy-typing error in the order with these values. The $E_{40}$ is absent in their work. 
\end{table}

\begin{table}[H]
\caption{\label{tab:table4}Comparison of eigenvalues of $V(r)=-\frac{1}{r}+0.01r$ for $l$ states.}
\begin{ruledtabular}
\begin{tabular}{cccccccc}
 $n$&$l$&$E_{0l}$& \cite{13} &\cite{32} \\
\hline
0& 0 & $-0.221030285825463$ & $−0.221030563404$ &$-0.221031$ \\
 & 1 & $0.0174005540419951$ & $0.017400552510$ &$0.0174006$ \\
 & 2 & $0.1024748170092927$ & $0.102472150415$ & $0.102472$\\
 & 3 & $0.1598337896864147$ & $0.159830894613$ & $0.159831$\\
 & 4 & $0.2062381776256258$ & $0.206238109687$& $0.206238$\\
 & 5 & $0.2466820720535486$ & $0.246682072100$ & $0.246681$\\
\end{tabular}
\end{ruledtabular}
\end{table}

It can be seen from Tables \ref{tab:table3} and \ref{tab:table4} that, eigenvalues of the radially and orbitally excited states by ANN method is in good agreement with the reference studies.

\subsection{Case \textrm{3}:  $V(r)=-\frac{1}{r}+100r$}
The Hamiltonian for this potential is
\begin{equation}
H=-\frac{1}{2} \frac{d^2}{dr^2}-\left(-\frac{1}{r}+100r- \frac{l(l+1)}{2r^2}   \right).
\end{equation}
In Tables \ref{tab:table5} and \ref{tab:table6}, we present eigenvalues of the potential $V(r)=-\frac{1}{r}+100r$ for $n$ and $l$ states, respectively.

\begin{table}[H]
\caption{\label{tab:table5}Comparison of eigenvalues of $V(r)=-\frac{1}{r}+100r$ for $n$ states.}
\begin{ruledtabular}
\begin{tabular}{cccccccc}
 $l$&$n$&$E_{n0}$& \cite{13} &\cite{32} \\
\hline
0& 0 & $46.402327405478400$ & $46.402258652779$ &$46.40221$\\
 & 1 & $85.339352424879541$  & $85.339271687574$ &$85.3393$\\
 & 2 & $116.72879854879542$  & $116.728692980119$ &$116.729$\\
 & 3 & $144.31559756874218$ & $144.315456241781$ &$144.315$\\
 & 4 & $169.46073150674210$ & $169.460543870657$ &$169.461$\\
 & 5 & $192.85053644685147$  & $192.850291861086$&$192.851$ \\

\end{tabular}
\end{ruledtabular}
\end{table}

\begin{table}[H]
\caption{\label{tab:table6}Comparison of eigenvalues $V(r)=-\frac{1}{r}+100r$ for $l$ states.}
\begin{ruledtabular}
\begin{tabular}{cccccccc}
 $n$&$l$&$E_{0l}$& \cite{13} &\cite{32} \\
\hline
0& 0 & $46.402327405478400$ & $46.402258652779$ &$46.4022$ \\
 & 1 & $70.018265184426244$ & $70.016058921076$ &$70.0161$ \\
 & 2 & $89.717285390446149$ & $89.715370910984$ & $89.7154$\\
 & 3 & $107.36655900853525$ & $107.334329106273$ & $107.334$\\
 & 4 & $123.56178707194739$ & $123.561985764157$& $123.562$\\
 & 5 & $138.7608054659801$ & $138.761138633388$ & $138.761$\\
\end{tabular}
\end{ruledtabular}
\end{table}

It can be seen from Tables \ref{tab:table5} and \ref{tab:table6} that, eigenvalues of the radially and orbitally excited states by ANN method is in good agreement with the reference studies. In this case, numerical results of orbitally excited sates can differ in first two or three digits. The reason for this difference can be the dominance of the linear part of the potential compared to the Coulomb part.

\subsection{Case \textrm{4}:  $V(r)=-\frac{a}{r}+r$}
The Hamiltonian for this potential is
\begin{equation}
H=-\frac{1}{2} \frac{d^2}{dr^2}-\left(-\frac{a}{r}+r- \frac{l(l+1)}{2r^2}   \right).
\end{equation}
In Table  \ref{tab:table7}, we present ground state eigenvalues of the potential $V(r)=-\frac{a}{r}+r$ for varying $a$ values. It can be seen from Table \ref{tab:table7} that obtained results are in good agreement with the given studies.

\begin{table*}
\caption{\label{tab:table7}Comparison of ground state eigenvalues for  $V(r)=-\frac{a}{r}+r$. }
\begin{ruledtabular}
\begin{tabular}{ccccccccc}
 $a$ & $E_{00}$ & \cite{1} & \cite{13}& $a$ & $E_{00}$  & \cite{9}& \cite{13}\\ \hline
0.2 & 2.16731633527814569 & 2.167316 & 2.167316208772717 & 0.1 & 2.25367805603652147 & 2.253678 & 2.253678098810761 \\

0.4 & 1.98850421340265894 & 1.988504 & 1.988503899750869 & 0.3 & 2.07894965124100145 & 2.078949 & 2.078949440194840 \\

0.6 & 1.80107438954743241 & 1.801074 & 1.801073805646947 & 0.5 & 1.89590541258401687 & 1.895904 & 1.895904238476994 \\

0.8 & 1.60440950560142422 & 1.604410 & 1.604408543236585 & 0.7 & 1.70393557501874021 & 1.703935 & 1.703934818031980 \\

1.0 & 1.39787578582763672 & 1.397877 & 1.397875641659907 & 0.9 & 1.50241669742698608 & 1.502415 & 1.502415495453739\\

1.2 & 1.18083812306874152 & 1.180836 & 1.180833939744787 & 1.1 & 1.29071042316348541 & 1.290709 & 1.290708615983606\\

1.4 & 0.95264360884520136 & 0.952644 & 0.952640495218560 & 1.3 & 1.06817386102041328 & 1.068171 & 1.068171244486971\\

1.6 & 0.71266200850041624 & 0.712662 & 0.712657680461034 & 1.5 & 0.83416589280625410 & 0.834162 & 0.834162211049953\\

1.8 & 0.46026600634169799 & 0.460266 & 0.460260113873608 & 1.7 & 0.58805423050569871 & 0.588049 & 0.588049168557953\\

\end{tabular}
\end{ruledtabular}
\end{table*}

\subsection{Heavy Quarkonium Spectra}
Quarkonium systems are an ideal area for clarifying our understanding of QCD. They probe nearly all the energy regimes of QCD from high energy region to low energy region. In the high energy region, an expansion of coupling constant is possible and perturbative QCD is applicable. In the low energy region, such an expansion is not possible in the coupling constant and therefore non-perturbative methods need to be used. Besides that non-relativistic QCD (NRQCD) approximation is also used for spectroscopy, decay and production of heavy quarkonium \cite{37,38}. For an overview of NRQCD, see \cite{39}.

In this section we obtained spin averaged mass spectra charmonium, bottomonium, and bottom-charmed system by solving nonrelativistic Schrödinger equation. It is possible to obtain full spectra for quarkonium systems including relativistic effects, spin-spin and spin-orbit interactions. Since most of these contributions are really small compared to the given potential, even by neglecting those effects one can find results that are close to the experimental data.

The related Cornell potential is \cite{40}
\begin{equation}
V(r)=Ar-\frac{\kappa}{r}
\end{equation}
with

\begin{eqnarray*}
A&=& 0.191 ~ \text{GeV}^2 \\
\kappa &=& 0.472  \\
m_b &=& 4.7485 ~ GeV \\
m_c &=& 1.3205 ~ GeV.
\end{eqnarray*}

In Tables \ref{tab:table8}, \ref{tab:table9} and \ref{tab:table10}, charmonium, bottomonium, and bottom-charmed spectra are presented, respectively.

\begin{table}[H]
\caption{\label{tab:table8}Charmonium spectra. All results are in GeV.}
\begin{ruledtabular}
\begin{tabular}{cccccccc}
State& This work & \cite{32} & Exp. \cite{41} \\
\hline
1S & $3.098$ & $3.096$ &$3.097$ \\
2S & $3.688$ & $3.672$ &$3.686$ \\
3S & $4.029$ & $4.085$ & $4.039$\\
1P & $3.516$ & $3.521$ & $3.511$\\
2P & $3.925$ & $3.951$ &$3.927$ \\
3P & $4.301$ & $4.310$ & -\\
1D & $3.779$ & $3.800$ & $3.770$\\
\end{tabular}
\end{ruledtabular}
\end{table}

\begin{table}[H]
\caption{\label{tab:table9}Bottomonium spectra. All results are in GeV.}
\begin{ruledtabular}
\begin{tabular}{cccccccc}
State& This work & \cite{32} & Exp. \cite{41} \\
\hline
1S & $9.460$ & $9.462$ &$9.460$ \\
2S & $10.026$ & $10.027$ &$10.023$ \\
3S & $10.354$ & $10.361$ & $10.355$\\
4S & $10.572$ & $10.624$ & $10.579$\\
1P & $9.891$ & $9.963$ & $9.899$\\
2P & $10.258$ & $10.299$ &$10.260$ \\
3P & $10.518$ & $10.564$ & $10.512$\\
1D & $10.156$ & $10.209$ & $10.164$\\
\end{tabular}
\end{ruledtabular}
\end{table}

\begin{table}[H]
\caption{\label{tab:table10}Bottom-Charmed spectra. All results are in GeV.}
\begin{ruledtabular}
\begin{tabular}{cccccccc}
State& This work & \cite{32} & Exp. \cite{41} \\
\hline
1S & $6.274$ & $6.362$ &$6.275$ \\
2S & $6.839$ & $6.911$ &$6.842$ \\
3S & $7.245$ & $7.284$ & \\
4S & $7.522$ & $7.593$ & \\
1P & $6.743$ & $6.792$ & \\
2P & $7.187$ & $7.178$ & \\
3P & $7.467$ & $7.494$ & \\
1D & $7.046$ & $7.051$ & \\
\end{tabular}
\end{ruledtabular}
\end{table}

As can be seen in Tables \ref{tab:table8}, \ref{tab:table9}, and \ref{tab:table10} charmonium, bottomonium, and bottom-charmed spectra of ANN agree well with given reference and available experimental data.

\section{\label{sec:level4}Summary}
In this paper, we applied ANN method to deal with the solution of the Schrödinger equation with Coulomb-plus-linear potential. This potential belongs to the non-solvable potentials class which have an exactly/analytically solvable part, together with a modifying term. We obtained the eigenvalues of Schrödinger equation and spin-averaged mass spectra of charmonium, bottomonium, and bottom-charmed systems. The obtained eigenvalues and heavy quarkonium spectra are in agreement with the theoretical studies and available experimental data.  

The feed forward ANNs method have a good property of function approximation. A function approximation problem is to select or find a function among a well defined functions set that closely matches a target function in a task specific way. This form employs a feed forward neural network as the basic approximation element, whose parameters (weights and biases) are adjusted to minimize an appropriate error function. In this study, the wave function is represented by the feed forward ANN and its inputs are taken as coordinate values. A trial solution is written as a feed forward neural network which contains adjustable parameters (weights and biases) and eigenvalue is refined to the known solutions by training the neural network. 

This study would be useful for the exact or quasi-exact spectra of a few body systems. It is also possible in principle  to handle many body problems but such problems will impose much more heavier computational load and other difficulties such as convergence of sigmoid functions \cite{36} and availability of hardware.

\newpage 

\end{document}